\documentclass[aps,prb,twocolumn,floatfix]{revtex4-1}

\pdfoutput=1

\usepackage{times}
\usepackage{amsfonts}
\usepackage{amssymb}
\usepackage{amsmath}
\usepackage{graphicx}
\usepackage{bm}
\usepackage{verbatim}

\usepackage{bm}
\usepackage{color}
\usepackage{stackrel}
\usepackage{accents}
\usepackage{hyperref}
\usepackage{latexsym}
\usepackage{float}

\usepackage{ulem}

\newcommand{\kk}{\mathbf{k}}
\newcommand{\gs}{\gamma_{\kk}^{s}}
\newcommand{\gn}{\gamma_{\kk}^{n}}

\newcommand{\gd}{\gamma_{\kk}^d}



\begin{document}

\title{
Interplay between nematicity and Bardasis-Schrieffer modes in the short-time dynamics of unconventional superconductors}

\author{Marvin A. M\"{u}ller$^1$, Pavel A. Volkov$^2$, Indranil Paul$^3$ and Ilya M. Eremin$^1$}
\affiliation{1- Institut f\"{u}r Theoretische Physik III, Ruhr-Universit\"{a}t Bochum, D-44801 Bochum, Germany \\
2 - Department of Physics and Astronomy, Center for Materials Theory, Rutgers University, Piscataway, New Jersey 08854, USA \\
3- Laboratoire Mat\'{e}riaux et Ph\'{e}nom\`{e}nes Quantiques, Universit\'{e} de Paris, CNRS, F-75013, Paris, France}

\begin{abstract}

Motivated by the recent experiments suggesting the importance of nematicity in the phase diagrams of iron-based and cuprate high-$T_c$ superconductors, we study the influence of nematicity on the collective modes inside the superconducting state in a non-equilibrium. In particular, we consider the signatures of collective modes in short-time dynamics of a system with competing nematic and $s$- and $d$-wave superconducting orders. In the rotationally symmetric state, we show that the Bardasis-Schrieffer mode, corresponding to the subdominant pairing, hybridizes with the nematic collective mode and merges into a single in-gap mode, with the mixing vanishing only close to the phase boundaries. For the d-wave ground state, we find that nematic interaction suppresses the damping of the collective oscillations in the short-time dynamics. Additionally, we find that even inside the nematic $s+d$-wave superconducting state, a Bardasis-Schrieffer-like mode leads to order parameter oscillations that strongly depend on the competition between the two pairing symmetries. We discuss the connection of our results to the recent pump-probe experiments on high-$T_c$ superconductors.

\end{abstract}

\maketitle

\section{Introduction}
The rapid development of pump-probe non-equilibrium spectroscopy opened new powerful tools to investigate collective excitations in condensed matter systems. In particular, probing relaxation dynamics in unconventional superconductors appears to be a promising field due to variety of different emerging phases accompanying superconducting ground state in these systems.\cite{Pashkin2010,Beck2011,Cav2011Science,Matsunaga2012,Conte2012,Beck2013,Mansart2013,Shimano2014Science,CavNature2014,Matsunaga2017,Katsumi2018,Nakamura2019,Chu2020} The dynamics of conventional charge superfluids is well understood by now. An intense pump pulse with a frequency of the order of the superconducting gap couples non-linearly to the Cooper-pairs, which leads to a coherent excitation of the superconducting order parameter $\Delta(t)$. It then performs a damped oscillation due to the existence of an intrinsic Higgs amplitude mode at $\omega_{\text{H}} = 2|\Delta(t=\infty)|$, which decays like $1/{\sqrt{|\Delta|t}}$\cite{Volkov1974,Amin2004,Barankov2004,Yuzbashyan2005,Yuzbashyan2006,Barankov2006,Papenkort2007,Krull2014,Big-Quench-Review2015,Tsuji2015,Aoki-Higgs2,Matt-Higgs,Cui2019,Schwarz2020,Mootz2020}. However, the rich phase diagram of unconventional superconductors with multiple competing states introduces further complexity. For example, an additional symmetry breaking, due to spin-\cite{Dzero2015} or charge-\cite{Moor2014,cea2014,Sentef2017} density wave instability or due to a competing superconducting state\cite{Foster2013,Mueller18,Kirmani2018, Mueller19} give rise to novel collective modes.  

One interesting example is an unconventional superconductor with a subdominant pairing interaction with a symmetry different from the ground state one. In that case a sharp collective Bardasis-Schrieffer mode (BS mode) associated with the 'failed' ground state is expected to be present within the superconducting gap. It has been originally introduced by Bardasis and Schrieffer in Ref. [\onlinecite{bardasis61}] for the subdominant $B_{1g}$ (d$_{x^2-y^2}$-wave)-symmetric mode in the $A_{1g}$ ($s$-wave)-symmetric ground state. Most recently, possible signatures of this BS mode were reported in the Raman response of the iron-based superconductors due to the close competition between $s_\pm$- and $d_{x^2-y^2}$-wave superconductivity in these compounds.\cite{maiti15,maiti16} At the same time, the situation in the iron-based and in several other unconventional superconductors is complicated by the presence of the nematic order and its fluctuations \cite{Lawler2010,cyr2015,Auvray2019,Murayama2019,Chu2010,Fradkin2010, Patz2014, Watson2016,Luo2017,Fernandes2019} in the normal state making further analysis on the origin of the anomalous enhancement of the $B_{1g} $ Raman signal\cite{Gallais2016,Boehm18} below superconducting transition temperature necessary. Most recently, we have analyzed the signatures of the BS mode in the short-time dynamics\cite{Mueller19}. This mode manifests itself as an additional undamped oscillation of the superconducting gap amplitude at $\omega_{\text{BS}} \leq \omega_{\text{H}}$ and its frequency depends on the strength of the residual interaction in the subdominant pairing symmetry channel. In addition its dependence on the fluence and polarization is distinct from the damped Higgs oscillations. At the same time, if the interaction in the subdominant symmetry channel is weak, i.e. the system is far from degeneracy point between two ground states, the Bardasis-Schrieffer mode becomes extremely close to the usual superconducting Higgs mode oscillation, making it challenging to distinguish the two.

In this manuscript we extend the analysis of collective modes in non-equilibrium superconductors \cite{Mueller19} to
include the effects of nematic order and its collective excitations. We address the question of whether pump-probe technique can be used to reveal an interplay between various collective modes visible in the superconducting state and to distinguish the Pomeranchuk nematic collective mode
from the BS mode due to the subdominant Cooper-pairing channel. In particular, we first calculate the short-time dynamics in the rotationally symmetric ground state, where we show that the nematic interaction softens the frequency of the BS mode, even when competition between different pairing symmetries is weak, but do not lead to the appearance of a second collective mode. This occurs due to a strong mixing between the collective nematic and BS mode at finite frequencies. Furthermore, for d-wave ground state the nematic interaction considerably improve the visibility of BS mode-induced oscillations. Second, we study the fate of the collective modes inside the nematic superconducting state and analyze their short-time dynamics, again finding only a single collective mode below the Higgs mode frequency. Finally, we discuss the consequences of our findings in the context of cuprate and iron-based high-$T_c$ superconductors.

\section{Model}
To model superconductivity in the presence of the nematicity and its collective modes, we consider the two-dimensional single-band model on the square lattice with nematic and superconducting mean-field order parameters. The nematic order, $\Delta_n \gamma^n_{\bf k}$ describes a Pomeranchuk-like instability and $\Delta_{\bf k}$ refers to the competing $s$-wave ($\Delta_\kk= \Delta_s\gamma_\kk^s$) and $d$-wave ($\Delta_\kk= \Delta_d\gamma_\kk^d$) superconducting order parameters or their mixture in the nematic and/or time-reversal symmetry broken state (${\Delta_\kk= \Delta_s\gamma_\kk^s + e^{i\alpha} \Delta_{d}\gamma_\kk^{d}}$) where $0\leq \alpha \leq\pi$. Here, $\Delta_s$ and $\Delta_d$ are corresponding magnitudes of the $s-$ and the $d-$wave superconducting order parameter, respectively. The Hamiltonian reads
\begin{align}\label{eq:ham}
    H = \sum_{\kk,\sigma} \left(\xi_\kk + \Delta_{n}\gn\right) c_{\kk\sigma}^\dagger  c_{\kk\sigma} + \sum_\kk \left[\Delta_\kk c_{\kk\uparrow}^\dagger c_{-\kk\downarrow}^\dagger + \text{h.c.}\right],
\end{align}
where $c_{\kk\sigma}^{(\dagger)}$ is the operator that annihilates (creates) an electron with spin $\sigma$ with momentum $\kk$. The function ${\xi_\kk = \alpha k^2 - \mu}$, with parameters $\alpha,\mu >0$, describes the band dispersion. We do not assume a continuous rotational symmetry, taking instead the parabolic form as an approximation for a full lattice tight-binding dispersion that has only discrete ($C_4$) rotational symmetry. Note that although we focus on $s$-wave ($A_{1g}$) and $d_{x^2-y^2}$-wave ($B_{1g}$) superconductivity, the considerations in this paper can be easily generalized to any superconducting order, which belongs to the even one-dimensional irreducible representations. For simplicity we will drop the index of $d_{x^2-y^2}$. We also choose $\gs = 1$ and $\gamma_\kk^{{d}} = \gamma_\kk^{{n}} = \sqrt{2}\cos(2\phi)$ similar to Ref. [\onlinecite{Chen20}]. \\
The conventional and helpful way to describe the short-time dynamics in the superconducting state\cite{Barankov2004,Yuzbashyan2005} is to introduce the Anderson pseudospin operators
\begin{align}\label{eq:pseudospins}
    \mathbf{s}_\kk = \frac{1}{2} \begin{pmatrix}
	c_{\kk\uparrow}^\dagger & c_{-\kk\downarrow}
\end{pmatrix}  \boldsymbol{\sigma}\begin{pmatrix}
	c_{\kk\uparrow} \\ 
	c_{-\kk\downarrow}^\dagger
\end{pmatrix} .
\end{align}
The operators $s_\kk^x,s_\kk^y$ and $s_\kk^z$ fulfill the spin commutation relations. This description can be generalized to the presence of the nematic order in a straightforward way. Using Anderson pseudospins Eq.~(\ref{eq:ham}) transforms into
\begin{align}
    H = \sum_\kk \mathbf{B}_\kk \cdot \mathbf{s}_\kk.
\end{align}
This form of the Hamiltonian describes a set of pseudospins $\mathbf{s}_\kk$ inside a pseudo-magnetic field ${\mathbf{B}_\kk = \left(2\Delta_\kk^\prime, 2\Delta_\kk^{\prime\prime},2\Tilde{\xi}_\kk\right)^T}$. Here we introduced the shorthand notation $\Delta_\kk = \Delta_\kk^\prime -i\Delta_\kk^{\prime\prime}$, where $\Delta_\kk^\prime$ and $-\Delta_\kk^{\prime\prime}$ are the real and imaginary part of $\Delta_\kk$, and the effective band structure $\tilde{\xi}_\kk = \xi_\kk + \Delta_n\gamma_\kk^n$. Note that the nematic order parameter $\Delta_n$ is real, while $\Delta_s$ and $\Delta_d$ can be complex numbers. Additionally, we introduce the pairing interactions $V_s$ and $V_d$ for the $s$- and $d$-wave pairing channel and the nematic interaction $V_n$, where $V_{s,d,n} < 0$ means attractive interaction in these channels. 
The order parameters need to be determined self-consistently via
\begin{figure}
    \centering
    \includegraphics[width=\linewidth]{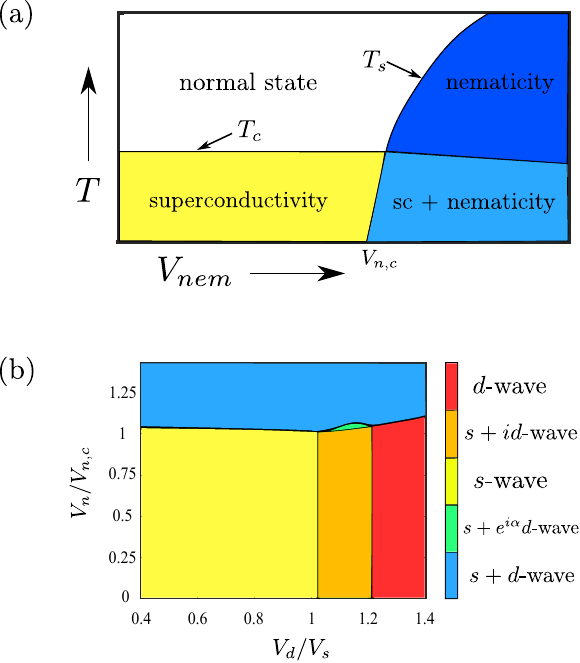}
    \caption{(a): Sketched phase diagram of the considered model. Above a critical value $V_{n,c}$ nematic order sets in, which leads to a mixed superconducting and nematic state below for $T<T_c$ and a pure nematic state for $T_c<T<T_s$. (b): Computed zero temperature phase diagram of the model, see Eq.~(\ref{eq:ham}). Solutions with finite nematic order $\Delta_n$ breaking rotational symmetry are characterized by coexisting $\Delta_s$ and $\Delta_d$ - order parameters for the s-wave and d-wave superconducting order, respectively. In the larger part of the phase diagram $\Delta_d$ and $\Delta_s$ can be chosen both real ($s$-,$d$-,or $s+d$-wave). However, once both channels are closely degenerate, an $s+id$-wave state, breaking time reversal symmetry, evolves, turning into an $s+e^{i\alpha}d$-wave state, once a finite but small $\Delta_n$ develops.     }
    \label{fig:phasediagram}
\end{figure}
\begin{align}\label{eq:self_n}
    \Delta_n &= V_n\sum_{\kk,\sigma}\gn\left\langle c_{\kk\sigma}^\dagger c_{\kk\sigma}\right\rangle = 2V_n\sum_\kk \gn \left \langle s_\kk^z\right\rangle,\\ \label{eq:self_s}
    \Delta_s &= V_s\sum_{\kk}\gs\left\langle c_{\kk\uparrow}^\dagger c_{-\kk\downarrow}^\dagger\right\rangle = V_s\sum_\kk \gs \left\langle s_\kk^x -i s_\kk^y\right\rangle,\\\label{eq:self_d}
    \Delta_d &= V_d\sum_{\kk}\gd\left\langle c_{\kk\uparrow}^\dagger c_{-\kk\downarrow}^\dagger\right\rangle = V_d\sum_\kk \gd \left\langle s_\kk^x -i s_\kk^y\right\rangle.
\end{align}
The nematic interaction needs to overcome a critical attractive strength, $V_{n,c} = -\frac{1}{2\nu_0} =  -2\pi\alpha$, where $\nu_0 = \frac{1}{4\pi\alpha}$ is the density of states, to form a nematic state at zero temperature \cite{Chen20}. Note that compared to Ref. [\onlinecite{Chen20}] our value for $V_{n,c}$ has an additional factor of 2, which stems from the explicit summation over spins in Eq.~(\ref{eq:self_n}). In contrast to that the superconducting instability is known to occur at an arbitrarily weak coupling, such that either $\Delta_s$ or $\Delta_d $ (or both) are always finite at $T=0$ as soon as the interaction in the corresponding channel is non-zero. Note that we vary the ratio $V_d/V_s$ and the parameter $V_n$ within this work and keep $V_s = 0.65V_{n,c}$ constant. In equilibrium the pseudospin expectation values are given by
\begin{align}
    \langle s_\kk^{x,eq}\rangle &= -\frac{\Delta_\kk^\prime}{2E_\kk}, \\
    \langle s_\kk^{y,eq} \rangle&= -\frac{\Delta_\kk^{\prime\prime}}{2E_\kk}, \\
    \langle s_\kk^{z,eq}\rangle &= -\frac{\tilde{\xi}_\kk}{2E_\kk},
\end{align}
where the quasiparticle energy dispersion ${E_\kk = \sqrt{\tilde{\xi}_\kk^2 + (\Delta_\kk^\prime)^2 + (\Delta_\kk^{\prime\prime})^2}}$.
The general interplay of superconductivity and the nematic order is sketched in Fig.\ref{fig:phasediagram}(a), while Fig.\ref{fig:phasediagram}(b) shows computed zero temperature phase diagram of this model. For $V_n< V_{n,c}$ one finds the phase diagram of competing $s$- and $d$-wave superconductivity; it consists of pure $s$-wave superconductivity below $V_d/V_s \approx 1$, pure $d$-wave superconductivity above approximately $V_d/V_s > 1.2$ and an $s+id$-wave state in between. In the $s+id$-wave state $\Delta_s$ and $\Delta_d$ have a phase-difference of $\pi/2$; this state breaks time-reversal symmetry, but is invariant under a combination of $C_4$-rotation symmetry and time-reversal symmetry. Around $V_n \approx V_{n,c}$ nematic order emerges, breaking rotational symmetry and turns the pure $s$- and pure $d$-wave states into mixed $s+d$-wave states. On the other hand, in the $s+id$-wave state, that has a very tiny region in parameter space, both $\Delta_s$ and $\Delta_d$ are nonzero even in the absence of nematic order. On increasing $V_n$, $s+id$-wave state evolves into the $s+d$-state via an intermediate $s+e^{i\alpha}d$ state \cite{kang2018}, that breaks both $C_4$ and time-reversal symmetry as well as their combination. Note that in the $s+d$-wave state we always find that the nematic order parameter $\Delta_n$ and the $d$-wave order parameter $\Delta_d$ have a different sign. 

We aim to understand the short-time dynamics of the order parameters in this model, which is closely connected to the short-time dynamics of the pseudospin expectation values as seen from Eqs (\ref{eq:self_n})-(\ref{eq:self_d}). Using the Heisenberg equation of motion one obtains Bloch-type equations for the time-evolution of the Anderson pseudospin operators
\begin{align}\label{eq:eom}
    \frac{d}{dt}\mathbf{s}_\kk = i\left[H,\mathbf{s}_\kk\right] =  \mathbf{B}_\kk\times \mathbf{s}_\kk.
\end{align}
\begin{figure}
    \centering
    \includegraphics[width=\linewidth]{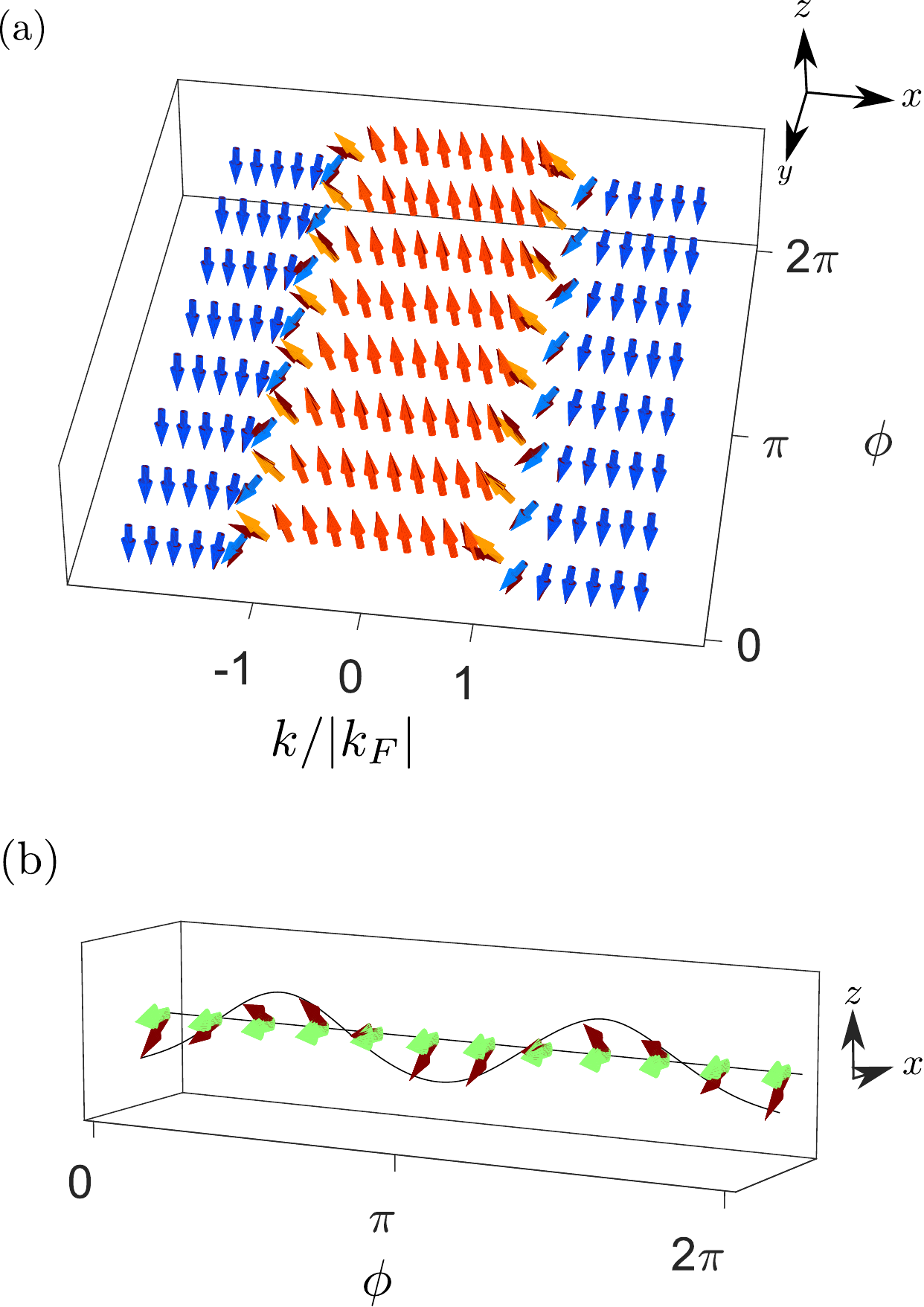}
    \caption{The direction of the pseudospins $s_{\kk}$ in the s-wave superconducting state (colored arrows) and the direction of the negative pseudomagnetic field $-\mathbf{B}_\kk$ after a quenched finite nematic order parameter $\Delta_n$ (black arrows) (a). The effect of the quench is particularly pronounced near the Fermi level, where the pseudospins change their direction. This effect is shown explicitly for $|\kk| = k_F$ (b), where the quenched nematic order parameter adds an additional modulation $\Delta B_{\kk}^z \sim \gn = \sqrt{2 }\cos(2\phi)$ to the pseudomagnetic field.\label{fig:pseudospins}}
\end{figure}
 The nematic order modifies the dispersion ${\xi_\kk \rightarrow \tilde{\xi}_\kk = \xi_\kk + \Delta_n\gn}$. The properties of the pseudospins for conventional $s$-wave superconductors have been discussed extensively in previous works\cite{spivak2004,Altshuler2005,Foster2013}. In the normal state the pseudospin expectation value points along $z$-axis and is given by $\langle s_\kk^{z,eq}\rangle = -0.5 \text{sgn} (\xi_\kk)$, i.e. we have a domain of spins pointing upwards for $|\kk| < k_F$, where $k_F$ is the Fermi momentum, and a domain of spins pointing downwards for $|\kk| > k_F$ and there is a hard domain wall at $|\kk| = k_F$. In the superconducting state the pseudospins gain a finite $x$- and $y$-component due to the opening of the superconducting gap, which leads to a continuous "twisting" from spin ups into spin downs around $k_F$ as shown in Fig.~\ref{fig:pseudospins}(a). In the equilibrium the pseudospins $\mathbf{s}_\kk$ point exactly antiparallel to a pseudomagnetic field $\mathbf{B}_\kk$ and thus, do not precess in the field as seen from Eq.~(\ref{eq:eom}). Quenching the system (independent of the quenched quantity) has always the same effect: The pseudomagnetic field $\mathbf{B}_\kk$ is tilted and, thus, the pseudospins $s_\kk$ are not pointing antiparallel to $\mathbf{B}_\kk$ anymore and start to precess.
 
To see how nematic order changes the picture we quench the nematic order to a finite value $\Delta_n = 0.1\mu$. This in turn induces a change in the $z$-component of $\mathbf{B}_\kk$, $B_\kk^z\rightarrow B_\kk^z + 2\Delta_n\gn$. The effect of this additional $\Delta B_\kk^z = 2\Delta_n\gn$ is strongest near $|\kk| \approx k_F$ as seen in Fig.~\ref{fig:pseudospins}(b). The inclusion of the effect of a time-dependent electromagnetic field via a time-dependent vector potential $\mathbf{A}(t)$ leads to a similar effect, as it enters the $z$-component of the pseudomagnetic field via a minimal coupling ${B_\kk^z = 2\tilde{\xi}_\kk \rightarrow \xi_{\kk + \mathbf{A}} + \xi_{\kk - \mathbf{A}}}$. The pseudomagnetic field $\mathbf{B}_\kk$ becomes tilted in $z$ direction, which leads to a precession of the pseudospins $\mathbf{s}_\kk$.

Analogous effect is also obtained if one quenches the superconducting order parameter. Quenching the order parameters $\Delta_s$ or $\Delta_d$ leads to a change in the components $B_\kk^x$ and $B_\kk^y$. This results also in a precession of the pseudospins $\mathbf{s}_\kk$. Thus, due to the three different order parameters there are multiple options how to quench the system and drive it out of the equilibrium.

\section{Short-time dynamics}
\begin{figure*}
    \centering
    \includegraphics[width=\textwidth]{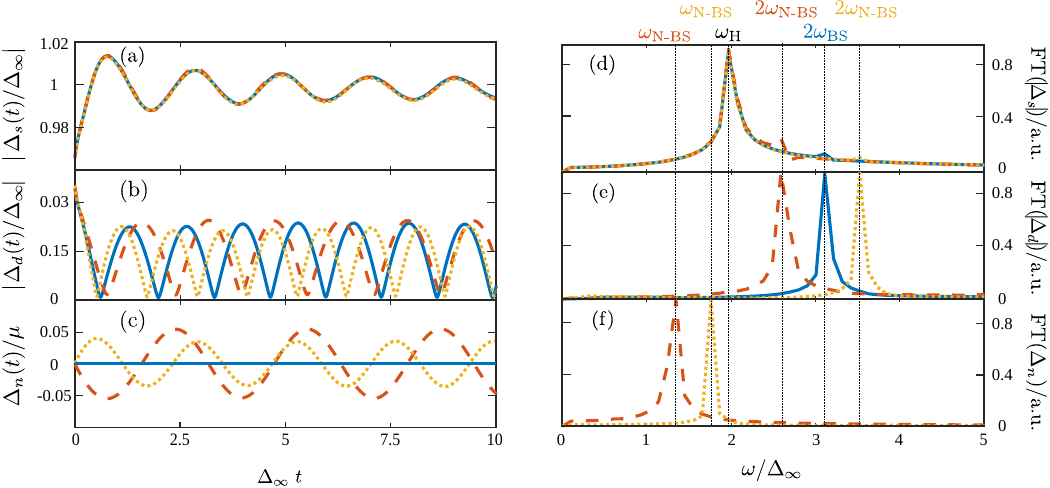}
    \caption{Short-time dynamics as follows from Eq.~(\ref{eq:eom}) after introducing the quench $\Delta_s\to0.95\Delta_s,\Delta_d = i 0.05 \Delta_s$. In all the cases the ground state is $s$-wave ($V_s/V_d=0.75$) but we consider three nematic interaction values, $V_n = 0$ (blue solid line), attractive $V_n = V_{n,c}/3$ (red dashed line) and repulsive $V_n = -V_{n,c}/3$ (yellow dotted line). The dynamics of the $s$-wave (a), $d$-wave (b) and nematic order $(c)$ are shown. The corresponding frequency spectra are shown on the right side (d)-(f).}
    \label{fig:BS_dynamics}
\end{figure*}
Using the equations of motion we are in the position to calculate the short-time dynamics for our model. In the following we discuss the short-time dynamics in the pure $s$-wave, pure $d$-wave and the mixed $s+d$-wave ground state for varying strength of the nematic interaction. 

In out-of-equilibrium experiments the ground state of these systems is excited by the application of a time-dependent laser field. This field pumps energy into the system and therefore excites Cooper pairs. However, in the non-nematic ground state, i.e., the pure $s$-wave or the pure $d$-wave state, the laser field does also temporarily break the rotational symmetry in these systems. This implies that even though one starts with a pure $s$-wave ($d$-wave) superconductor the pulse also excites $d$-wave ($s$-wave) Cooper pairs\cite{Mueller19}, although this pairing symmetry is energetically not favored and subdominant.

At the same time, the presence of the nematic order can also mix the $s-$ and $d-$ channels, which raises the question of the influence of nematic collective modes on the BS mode in a non-equilibrium system. In the following section we study the interplay of the BS and nematic collective modes in the short-time dynamics of our model.

\subsection{Superconducting $s$-wave ground state}\label{sec:s}

We first solve the equations of motion Eq.~(\ref{eq:eom}) inside the pure $s$-wave superconducting ground state and keep the superconducting $d$-wave state and the nematic state energetically nearby but subdominant. This resembles a simplified picture of the iron-based superconductors, which also show a strong presence of the $B_{1g}$ nematic\cite{Fernandes2014} and superconducting $d$-wave fluctuations\cite{Boehm18} nearby. We solve the self-consistency equations for $\Delta_s$ and quench its value by reducing it by a factor 0.95 and set the $d$-wave order parameter to $|\Delta_d| = 0.05|\Delta_s|$. The frequencies of the modes and the relative phases of oscillations have been found not to depend on the quench phase of $\Delta_d$. This perturbation is similar to the application of a time-dependent vector potential as discussed above. Approximating the vector potential pulse with a quench neglects possible polarization-dependent effects\cite{Mueller19} but makes the computation far more efficient. The resulting short-time dynamics is shown in Fig.~\ref{fig:BS_dynamics} for the ratio $V_d/V_s = 0.75$ and for different values of the nematic interaction strength $V_n$. Assuming no nematic interaction ($V_n = 0$) we find that the excited superconducting $s$-wave condensate is mainly oscillating with the Higgs-mode frequency $\omega_{\text{H}} = 2\Delta_{s,\infty}$. This oscillation with the Higgs-mode frequency shows typical $1/\sqrt{t}$ damping as its frequency is located at the border of the quasiparticle continuum\cite{Volkov1974}. At the same time, we also excite $d$-wave Cooper pairs due to the rotational symmetry breaking in the quench, which couple to $V_d$ and, thus, yields a finite order parameter $\Delta_d$. This $d$-wave condensate performs an undamped oscillation with the Bardasis-Schrieffer mode frequency $\omega_{\text{BS}} < \omega_{\text{H}}$ as discussed by us previously in Ref. [\onlinecite{Mueller19}]. Note that in {Fig.~\ref{fig:BS_dynamics} (b)} we show the absolute value of the complex order parameter $\Delta_d$, which oscillates around $0$ and thus, the frequency spectrum is peaked at $2\omega_{\text{BS}}$. Next we turn on a weak, attractive nematic interaction $V_n =  V_{n,c}/3 <0$ and find that the frequency $\omega_{\text{BS}}$ is immediately affected. In particular, it shifts to lower frequencies and its position is now determined by both $V_d $ and $V_n$. Additionally a finite nematic order parameter $\Delta_n$ is generated out of equilibrium, which magnitude oscillates around zero. It is important to note that the nematic order parameter does not oscillate with its own frequency but oscillates with the same frequency, i.e. by the modified $\omega_{\text{BS}}$, similar to $\Delta_d$. This occurs because the form factor of the $d-$wave superconducting order and that of the nematic order are the same, allowing both to mix (i.e. couple linearly) in the superconducting state. Thus, we find that the nematic fluctuations and the subdominant pairing fluctuations in the same $B_{1g}$ channel couple to form a single collective nematic Bardasis-Schrieffer mode, which we denote as $\omega_{\text{N-BS}}$ in the following. The nematic order parameter and the subdominant pairing channel perform undamped oscillations because their frequency is well below $\omega_{\text{H}} = 2\Delta_{s,\infty}$. At the same time, the nematic order parameter $|\Delta_n(t)|$ and the superconducting one, $|\Delta_d|$, show an antiphase oscillation, which means that although nematicity favors the $s+d$-wave state instead of the pure $s$-wave state, it still competes with superconductivity for the available phase space. To check the direct influence of the nematic interaction on the nematic Bardasis-Schrieffer mode we also considered repulsive nematic interaction as shown in Fig.~\ref{fig:BS_dynamics} for $V_n = -V_{n,c}/3$. We find that the frequency of $\omega_{\text{N-BS}}$ is pushed now to higher frequencies, as is expected from nematic/$d$-wave competition, and there is no additional excitation of a nematic mode seperated from the BS mode besides the nematic Bardasis-Schrieffer mode.

To get further insight in the interplay of the subdominant pairing instability and nematic fluctuations onto the collective modes in a superconducting system, we compute the frequency dependence of these modes within linear response theory. 
Our starting point are the equations of motion in Eq.~\ref{eq:eom}, which are linearized around their equilibrium values, i.e. we approximate $\mathbf{s}_\kk(t) \simeq \mathbf{s}_\kk^{eq} + \delta \mathbf{s}_\kk(t)$ and $\mathbf{B}_\kk(t) \simeq \mathbf{B}_\kk^{eq} + \delta \mathbf{B}_\kk(t)$. Thus, we obtain
\begin{align}\label{eq:lin_eom}
    \frac{d}{dt}\delta \mathbf{s}_\kk = \mathbf{B}_\kk^{eq}\times \delta \mathbf{s}_\kk+ \delta \mathbf{B}_\kk \times \mathbf{s}_\kk^{eq},
\end{align}
The perturbed pseudomagnetic field is determined by the perturbed order parameters
\begin{align}
\delta\mathbf{B}_\kk(t) = \left(2\delta\Delta_\kk^\prime(t),  2\delta\Delta_\kk^{\prime\prime}(t),2\delta\tilde{\xi}_\kk\right)^T\end{align} with $\delta\Delta_\kk(t) = \delta\Delta_s(t)\gs + \delta\Delta_d(t)\gd$ and $\delta\tilde{\xi}_\kk \simeq \frac{1}{2}\frac{\partial^2\xi_\kk}{\partial k^2}\mathbf{A}^2(t) + \delta\Delta_n(t)\gn = \alpha \mathbf{A}^2(t)  + \delta\Delta_n(t)\gn$. The linearized equations of motion in Eq.~(\ref{eq:lin_eom}) can be solved via a Fourier-transformation, which then gives 
\begin{align}\label{eq:sMB}
    \delta\mathbf{s}_\kk(\omega) = M_\kk(\omega) \delta\mathbf{B}_\kk(\omega),
\end{align}
where $M_\kk(\omega)$ is a matrix 
\begin{align}\label{eq:M}
    &M_\kk = \frac{1}{2E_\kk\left(\omega^2-4E_\kk^2\right)}\cdot \nonumber\\
    &\begin{pmatrix}
	2((\tilde{\xi}_\kk)^2 + (\Delta^{\prime\prime}_\kk)^2) & -2\Delta_\kk^\prime\Delta_\kk^{\prime\prime} + i\tilde{\xi}_\kk\omega & -2\tilde{\xi}_\kk\Delta_\kk^{\prime} - i\Delta^{\prime\prime}_\kk\omega \\ 
	-2\Delta_\kk^\prime\Delta_\kk^{\prime\prime} - i\tilde{\xi}_\kk\omega & 2((\tilde{\xi}_\kk)^2 + (\Delta^{\prime}_\kk)^2) & -2\tilde{\xi}_\kk\Delta_\kk^{\prime\prime} + i\Delta^\prime_\kk\omega \\ 
	-2\tilde{\xi}_\kk\Delta_\kk^{\prime} + i\Delta^{\prime\prime}_\kk\omega & -2\tilde{\xi}_\kk\Delta_\kk^{\prime\prime} - i\Delta^{\prime}_\kk\omega & 2((\Delta^{\prime}_\kk)^2+ (\Delta^{\prime\prime}_\kk)^2)
\end{pmatrix} 
\end{align}
Note that all order parameters in the matrix $M_\kk$ are at equilibrium. The perturbed pseudomagnetic field $\delta \mathbf{B}_\kk$ contains both, the information on the  $\delta\Delta_\kk$ and $\delta\Delta_n$, as well as on the vector potential $A$. It can be rewritten in terms of these quantities as follows
\begin{align}
    \delta\mathbf{B}_\kk &= 2\left(\delta\Delta_\kk^\prime,\delta\Delta_\kk^{\prime\prime},\alpha A^2(t) + \delta\Delta_n\gn\right)^T \\&= 2 G_\kk \left(\delta\Delta_s^\prime,\delta\Delta_s^{\prime\prime},\delta\Delta_d^\prime,\delta\Delta_d^{\prime\prime},\delta\Delta_n\right)^T  + 2\alpha A^2\hat{e}_z \\&\equiv 2G_{\kk}\delta\boldsymbol{\Delta} + 2\alpha A^2\hat{e}_z.
\end{align}
Here we introduced the matrix
\begin{align}
    G_\kk = \begin{pmatrix}
	\gs & 0 & \gd & 0 & 0 \\ 
	0 & \gs & 0 & \gd & 0 \\ 
	0 & 0 & 0 & 0 & \gn
\end{pmatrix},
\end{align}
which contains all relevant form factors. Using Eqs. (\ref{eq:self_n})-(\ref{eq:self_d}) and $V = \text{diag}(V_s,V_s,V_d,V_d,2V_n)$we find
\begin{align} \label{eq:gap_pseudo}
    \delta\boldsymbol{\Delta} = \sum_\kk V G_\kk^T\delta \mathbf{s}_\kk.
\end{align}
Inserting Eq.~(\ref{eq:sMB}) into Eq.~(\ref{eq:gap_pseudo}) we obtain 
\begin{align}\label{eq:delta_w}
\delta \boldsymbol{\Delta}(\omega)  = \left(\mathrm{1} - \chi(\omega)\right)^{-1} (2\alpha \mathcal{F}(A^2)(\omega))\sum_\kk V G_\kk^T M_\kk \hat{e}_z
\end{align}
with $\chi(\omega) = 2\sum_\kk V G_\kk^T M_\kk G_\kk$ and $\mathcal{F}(A^2)(\omega)$ is the Fourier-transformation of $A^2(t)$. The $5\times5$ matrix $\chi(\omega)$ can be understood as the response function of the gaps $\delta\boldsymbol{\Delta} = \left(\Delta_s^\prime,\Delta_s^{\prime\prime},\Delta_d^\prime,\Delta_d^{\prime\prime},\Delta_n\right)^T$. In the simplest case the resonance frequencies of $\delta\Delta(\omega)$ are given at those values of $\omega$, where the r.h.s. of Eq.~(\ref{eq:delta_w}) diverges. This is the case, when the condition
\begin{align}\label{eq:lin_omega}
    \det(\mathrm{1}-\chi(\omega))= 0
\end{align}
is fulfilled. As is expected in linear response, the form of the vector potential does not affect the frequencies.

Without nematic interaction Eq.~(\ref{eq:lin_omega}) reveals three real solutions: (i) the phase fluctuation (Goldstone mode) at $\omega = 0$, which becomes gapped due to the coupling to the fluctuation of the vector potential and is shifted inside the quasiparticle continuum\cite{Anderson1963}; (ii) the amplitude fluctuation (Higgs mode) at $\omega_{\text{H}} = 2\Delta_{\text{max}}$, of the dominant superconductor order parameter; and (iii) the Bardasis-Schrieffer mode, which occur due to residual attractive interaction in the competing superconducting channel. The frequency position of the Bardasis-Schrieffer mode is well separated from the Higgs-mode oscillation provided the system is sufficiently close to the secondary instability, i.e. $V_d$ is close to $V_s$.

\begin{figure}
    \centering
    \includegraphics[width=\linewidth]{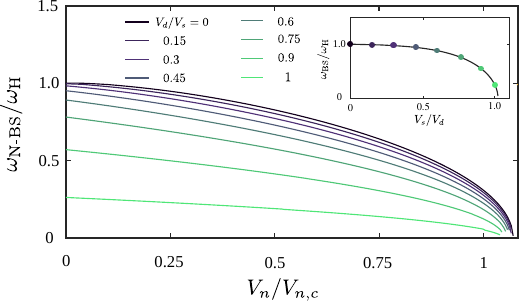}
    \caption{The dependence of the nematic Bardasis-Schrieffer mode frequency $\omega_{\text{N-BS}}$ with respect to the nematic interaction $V_n$ is computed from Eq.~(\ref{eq:lin_omega}). For various ratios of $V_d/V_s$ we track $\omega_{\text{BS}}$ until it drops to $0$. The inset shows the dependence of the pure Bardasis-Schrieffer mode $\omega_{\text{BS}}$ on $V_d/V_s$ without the nematic interactions}.
    \label{fig:enhanced_BS}
\end{figure}

Upon inclusion of the attractive nematic interaction this picture changes as we have observed in the short-time dynamics in Fig.~\ref{fig:BS_dynamics}. We find that the frequency of the Bardasis-Schrieffer mode $\omega_{\text{BS}}$ decreases upon increasing nematic interaction in the same $d$-wave channel and reaches zero at the phase transition where the nematic ground state forms. In particular, in Fig.~\ref{fig:enhanced_BS} we show the evolution of the nematic Bardasis-Schrieffer mode frequency upon increasing nematic interaction. Note that the critical $V_{n,c}$ at which a nematic ground state form varies only marginally with increasing $V_d/V_s$. Most interestingly, we find that this mode transforms into pure nematic mode for the case $V_d/V_s = 0 $. This indicates that collective BS-like mode in the superconducting state can be easily mixed with a purely nematic mode. This coupling of the nematic mode with the BS mode is also the present for repulsive nematic interactions, as discussed in Appendix \ref{app:B}.

We can further demonstrate this analytically. Assuming a pure s-wave groundstate we can calculate the response function $\chi(\omega) = 2\sum_\kk V G_\kk^T M_\kk G_\kk$. The sum over momenta $k$ can we rewritten as an integral over the angle $\phi$ and an integral over $\xi$ times the density of states $\nu_0$. This yields (the terms that vanish after $\xi$ or $\phi$ integral are omitted)
\begin{align}
&\chi(\omega) = \nu_0\int d\xi \frac{1}{\sqrt{\xi^2+\Delta^2}\left(\omega^2-4(\xi^2+\Delta^2)\right)}\cdot \nonumber \\
&\begin{pmatrix}
2V_s\xi^2 & 0 & 0 & 0 & 0 \\ 
0 & 2V_s(\Delta^2+\xi^2) & 0 & 0 & 0 \\ 
0 & 0 & 2V_d\xi^2 & 0 & 0 \\ 
0 & 0 & 0 & 2V_d(\xi^2+\Delta^2) & iV_d\omega\Delta \\ 
0 & 0 & 0 & -2iV_n\omega\Delta & 4V_n\Delta^2
\end{pmatrix}.
\end{align} 
One finds that the dynamics of $\Delta''_d$ and $\Delta_n$ decouple from that of $\Delta_s', \Delta_s'', \Delta_d$. Note that the first three entries on the diagonal of $\chi(\omega)$ correspond to the decoupled response of $\Delta_s^\prime$, $\Delta_s^{\prime\prime}$, $\Delta_d^\prime$, while the response of $\Delta_d^{\prime\prime}$ and $\Delta_n$ appears to be coupled.
Plugging this into our condition for collective modes in Eq.~(\ref{eq:lin_omega}) we obtain 

\begin{align}\label{eq:det}
\det(1-\chi(\omega)) =&\nonumber \\ (1+2V_sI_1(\omega))&(1+2V_s(I_1(\omega)+I_2(\omega)))(1+2V_dI_1(\omega))\nonumber \\
\cdot	\Bigg[
	(1+&2V_d(I_1(\omega)+I_2(\omega))(1+4V_nI_2(\omega))  \nonumber\\&- 2V_nV_d\frac{\omega^2}{\Delta^2}I_2(\omega))^2
	\Bigg],
\end{align}
where we introduced the integrals
\begin{align}
	I_1(\omega) &= \nu_0 \int d\xi \frac{\xi^2}{\sqrt{\xi^2+\Delta^2}\left(4(\xi^2+\Delta^2)-\omega^2\right)}, \\
	I_2(\omega) &= \nu_0 \int d\xi \frac{\Delta^2}{\sqrt{\xi^2+\Delta^2}\left(4(\xi^2+\Delta^2)-\omega^2\right)}
\end{align}
We want to discuss each part of the expression in Eq.~(\ref{eq:det}). Note that the integrals $I_1(\omega)$ and $I_2(\omega)$ are monotonically increasing. Therefore the expressions $1+2V_sI_1(\omega)$, $1+2V_s(I_1(\omega) + I_2(\omega))$ and $1+2V_dI_1(\omega)$ can only have a single root at maximum and thus each term can give rise to a only single collective mode. The first term corresponds to the Higgs-mode ($\omega=2\Delta$):
\begin{align}
	1+2V_sI_1(\omega=2\Delta) = 1+ V_s\nu_0 \int d\xi \frac{1}{2 \sqrt{\xi^2+\Delta^2}} = 0
\end{align}
This expression is zero because it is the exact condition of the self-consistency equation for $\Delta$. The second term would correspond to the Goldstone-mode (here it is  at $\omega=0$ but becomes gapped upon inclusion of the gauge fluctuations):
\begin{align}
&1+2V_s(I_1(\omega=0) + I_2(\omega=0)) \nonumber\\ &= 1+ V_s\nu_0 \int d\xi \frac{1}{2 \sqrt{\xi^2+\Delta^2}} = 0,
\end{align}
which is again the self-consistency equation for $\Delta$. Since $I_1(\omega)$ is monotonically increasing and $|V_d|<|V_s|$ the part $1+2V_dI_1(\omega)$ has no root at all for $\omega \in [0,2\Delta]$. Therefore the coupled nematic Bardasis-Schrieffer mode corresponds to a root in the last term emerging from the coupling of $\Delta_d^{\prime\prime}$ and $\Delta_n$. 
Separating the response of $\Delta_d''$ and $\Delta_n$ in Eq.~\ref{eq:det} from the rest we can write a condition for the nematic Bardasis-Schrieffer mode $\omega_{\text{N-BS}}$ as
\begin{align}
&\det(\chi_{\text{N-BS}}(\omega)-1)=0, \label{eq:matrix_chidet} \\
\chi_{\text{N-BS}}(\omega) &= \begin{pmatrix} -2V_d(I_1(\omega)+I_2(\omega)) & iV_d\omega I_2(\omega)/\Delta \\
 -2iV_n\omega I_2(\omega)/\Delta & -4V_nI_2(\omega)
 \end{pmatrix}.\label{eq:matrix_chi1}
\end{align}

One observes, that there is a cross term in Eq.~(\ref{eq:matrix_chi1}) of the from $\sim i\omega \Delta_d''\Delta_n$ corresponding to a dynamic coupling between the nematic and the imaginary part of the d-wave order parameter. Indeed, such a coupling is allowed by symmetry, since $\Delta_d''$ breaks time reversal symmetry and both break rotational symmetry. On purely symmetry grounds, one may also expect a nonzero coupling between $\Delta_d'$ and $\Delta_n$; however it vanishes after integration over $\xi$, which comes as a result of linearization of the electronic spectrum. Thus, if the linearization of the spectrum around Fermi surface is not used, this coupling would be nonzero, although it is expected to be small, coming from the region away from the Fermi surface.

Using the self-consistency equation one gets ${I_1(\omega)=-\frac{1}{2 V_s}-\left(1-\frac{\omega^2}{4\Delta^2}\right)I_2(\omega)}$; $I_2(\omega)$ can be evaluated analytically. Then, Eq. \eqref{eq:matrix_chidet} reduces to the following form:
\begin{equation}
-\nu_0 V_s \frac{\arccos \sqrt{1-x^2}}{ x\sqrt{1-x^2}} 
=
\frac{
	1
}
{
v_n+\frac{v_d}{1-v_d}x^2	
},
\label{eq:BSnemLin}
\end{equation}
where $x=\frac{\omega}{2\Delta}$ and the notation  $v_{d}=V_{d}/V_s$ and $v_n = 2V_n/V_s$ has been introduced. L.h.s. of the Eq.~\eqref{eq:BSnemLin} is a monotonically increasing function, while r.h.s. is a decreasing one of $x$. Consequently, only one solution of the equations is possible; moreover, r.h.s. diverges at $x=1$, so a solution is guaranteed to exist up until the mode becomes unstable. An instability (solution at $x=0$) occurs either for $v_n = \frac{1}{\nu_0 |V_s|}\to V_n = -\frac{1}{ 2\nu_0} = V_{n,c}$ (independent of $v_d$) or for $v_d=1$ (independent of $v_n$). All of this is in agreement with the results in phase diagram in Fig.~\ref{fig:phasediagram}, also consistent with the absence of nematic/BS mixing at $\omega=0$. Most importantly, the mixing between two modes is clearly responsible for this "one-mode" behavior.

Thus, we come to the conclusion that in an $s$-wave ground state with competing $d$-wave and nematic states, there is a single collective in-gap mode that generically has a mixed character, although the mixing is weaker near an instability, where the frequency of the excitation goes to zero. This is evident from Fig.~\ref{fig:enhanced_BS}, where on decreasing $\omega_{\text{N-BS}}$, its dependence on $V_n$ becomes profoundly weaker. On the other hand, we find the coupling between $\Delta_d'$ and $\Delta_n$ to vanish after integration over the radial direction, as it is proportional to $\xi_k$.

We find that in the presence of nematic ordering tendency, in-gap collective mode cannot be immediately related to the competing superconducting correlations in the $d$-wave channel as was originally proposed\cite{Boehm18}, unless extremely close to a phase boundary, where the criticality of one order dominates. Moreover, close to a nematic phase, an oscillation of the superconducting order parameter can be generated by the interaction originating in the particle-hole nematic channel even for $V_d/V_s = 0$. We also emphasize that we do not observe two separate modes corresponding to the competition between the subdominant $d$-wave ground state and the nematic interaction because both are in the $B_{1g}$ channel. This changes if the nematic order has, for example, $B_{2g}$-type symmetry, where the upper found dynamics coupling vanishes as shown in Appendix \ref{app:C}. 

\subsection{Superconducting $d$-wave ground state}
\begin{figure*}
    \centering
    \includegraphics[width=\textwidth]{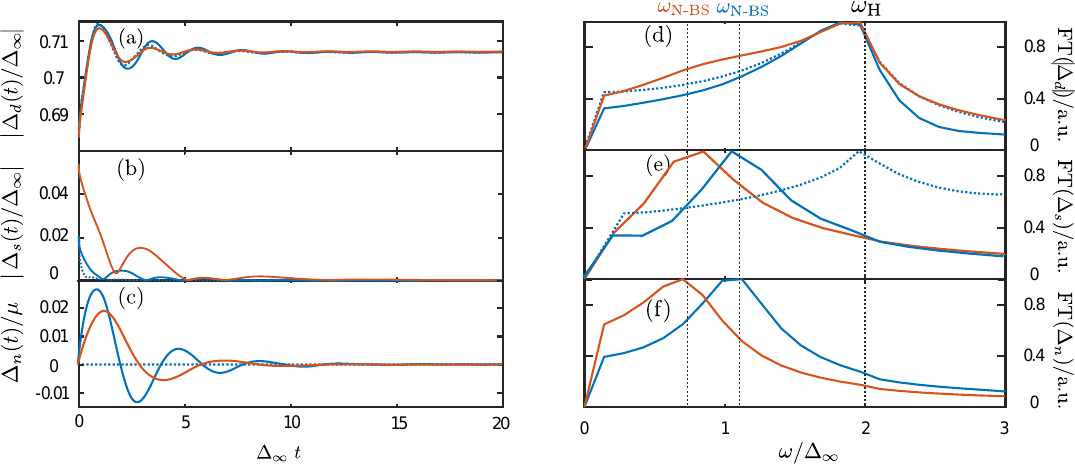}
    \caption{Short-time dynamics computed from Eq.~(\ref{eq:eom}) inside the $d$-wave ground state after the quench. Here we consider $V_n = 2V_{n,c}/3$ and vary the strength of the $s$-wave interaction channel $V_s = 0.25V_d$ (blue) and $V_s = 0.65V_d$ (red). Additionally we show the dynamics for vanishing nematic interaction $V_n=0$ for $V_s = 0.25V_d$ (dotted blue). On the right side, the dynamics of the induced $s$-wave order parameter $|\Delta_s|(t)$ (a), the $d$-wave order parameter $|\Delta_d(t)|$ (b) and the nematic order parameter $\Delta_n(t)$ (c) are shown. The corresponding frequency spectra are shown on the right side (d)-(f). }
    \label{fig:BSdwave_dynamics}
\end{figure*}
Now we turn to the discussion of the short-time dynamics in the pure $d$-wave superconducting ground state, which is present if $V_d/V_s > 1.2$ and for weak nematic interactions $V_n < V_{n,c}$. This scenario applies for example to the high-$T_c$ cuprate superconductors, where the presence of nematicity and nematic fluctuations is intensively discussed\cite{Lawler2010,cyr2015,Auvray2019,Murayama2019}. However, unlike in iron-based superconductors no strong competition between the dominant $d$-wave superconducting channel and a subdominant $s$-wave Cooper pairing channel is expected. Note also that usual BS modes from the secondary Cooper pairing instabilities are strongly damped due to the nodal structure of the superconducting gap\cite{Mueller19}. Therefore, nematic fluctuations in the $B_{1g}$ channel would be the only source of the visible additional oscillations in the superconducting order parameter.

We solve again the self-consistency equations and quench this time the finite $\Delta_d$ by a factor 0.95 and set the $s$-wave order parameter $|\Delta_s|=0.05|\Delta_d|$ similar to the previous subsection. In Fig.~\ref{fig:BSdwave_dynamics} we show the resulting short-time dynamics. Due to the strong damping of the collective dynamics the frequency spectra are much broader than in the $s$-wave ground state. We find that the $d$-wave order parameter mainly oscillates with its Higgs-mode frequency $\omega_{\text{H}} = 2\Delta_{\text{max}}$.
Additionally we see that the induced $s$-wave Cooper pairs couple to $V_s$ and generate a finite $\Delta_s$, which performs strongly damped oscillation, making its actual experimental observation difficult. However, we observe that in the presence of attractive nematic interaction, an additional oscillation frequency below $\omega_{\text{H}}$ can be resolved.

As shown in Fig.~\ref{fig:BSdwave_dynamics} this frequency is not a pure nematic mode depending only on the strength of $V_n$, but indeed depends on the strength of the attractive $s$-wave interaction, 
corresponding to the mixed nematic Bardasis-Schrieffer mode discussed before. If the residual attractive interaction in the nematic channel is strong, the frequency of this mode is well below $\omega_{\text{H}}$. We attribute the observed influence of the nematic interaction on the BS mode to the  form-factor of the nematic order parameter, that resembles that of the $d-$wave superconducting order parameter, having minima in the nodal region, where most quasiparticles are excited. Consequently, one expects damping to be weaker for the collective oscillations of the nematic order. However, due to the coupling between the BS and nematic mode, only a single mixed collective mode is observed. Thus, our results show that a not too strong nematic interaction effectively makes the BS mode less sensitive to damping by the nodal quasiparticles, making its observation in experiments more likely.

It is tempting to identify this nematic Bardasis-Schrieffer mode with the additional oscillation found recently in high-T$_c$ cuprates\cite{Chu2020}. Note that the nematic fluctuations has been found in the entire doping range of the hole-doped cuprates but specifically were related to the pseudogap region\cite{Murayama2019,Auvray2019}.

\subsection{Superconducting $s+d$-wave ground state}\label{sec:s+d}
\begin{figure*}
    \centering
    \includegraphics[width=\textwidth]{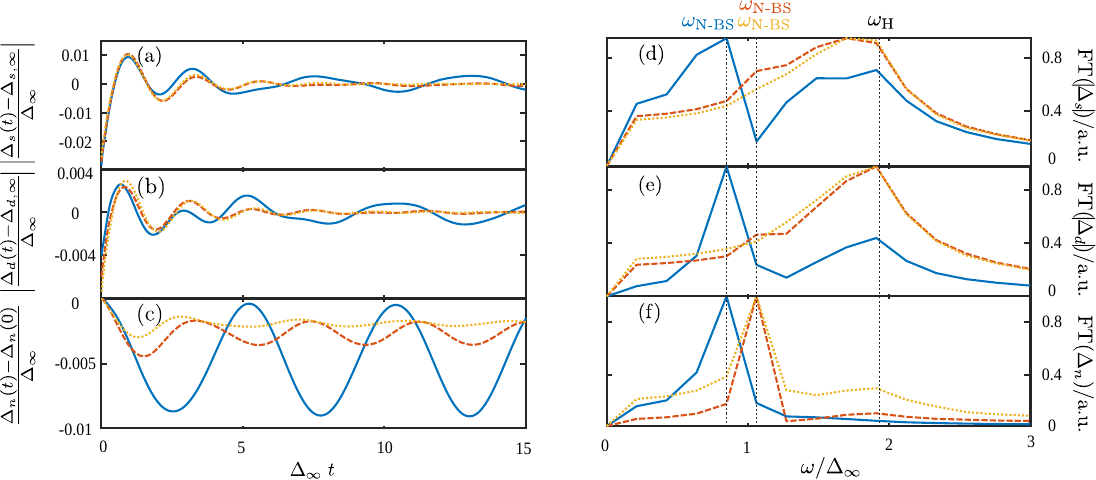}
    \caption{ Short-time dynamics in the nematic $s+d$-wave state as obtained from the solution of Eq.~(\ref{eq:eom}) after introducing quenches. Here we started in all cases with an $s+d$-wave ground state and quenched the superconducting order parameter by 5\% with respect to the equilibrium, i.e. $\Delta_s = 0.95\Delta_{s,eq}$ and $\Delta_d=0.95\Delta_{s,eq}$. We considered $V_d = 0.75V_s$ and three different values for the nematic interaction, $V_n = 1.1V_{n,c}$ (blue solid line), $V_n =  1.2V_{n,c}$ (red dashed line) and $V_n = 1.3V_{nc}$ (yellow dotted line). The dynamics of the $s$-wave (a), $d$-wave (b) and nematic order $(c)$ are shown. The corresponding frequency spectra are shown on the right side (d)-(f).}
    \label{fig:BS+nem_shorttime}
\end{figure*}
If the nematic interaction exceeds the critical value $V_{n,c}$ a nematic ground state forms and the tetragonal $C_4$ symmetry is broken. This breaking of the point group symmetry leads to a mixing of irreducible representations, with implications for superconducting state. In particular, a mixed two-fold symmetric composite $s+d$-wave state forms, where $s$- and $d$-wave order parameters mutually coexist. Nevertheless, as both $s$-wave and $d$-wave orders are driven by the separate interactions, one can still expect to distinguish a corresponding oscillations of their constituents in short time dynamics  and separate the main (dominant) pairing instability and a secondary (subdominant) one. Additionally, an amplitude mode of the nematic order can be expected, expecially close to $V_{n,c}$ To study this regime, we set $V_n > V_{n,c}$ and calculate the superconducting order parameters $\Delta_s$ and $\Delta_d$ and the nematic order parameter $\Delta_n$ self consistently. We quench both superconducting order parameters by a factor 0.95. The resulting short time dynamics is shown in Fig.~\ref{fig:BS+nem_shorttime}. One finds that both magnitudes, namely the $s$-wave and $d$-wave order parameters oscillate with two frequencies. The first one corresponds to the Higgs-mode $\omega_{\text{H}} = 2\Delta_{\text{max}}$ and a second mode with a lower frequency, which depends on the strength of the secondary (weaker) instability and the nematic interaction. In addition, while Higgs oscillations are damped as expected for an anisotropic superconductors as the gap maxima still lie within a quasiparticle continuum, the second mode, which we identify again as the nematic Bardasis-Schrieffer mode, can be strongly damped or undamped depending on whether the frequency of this mode is larger or smaller than  $2\min{\Delta_\kk}$. The latter depends on the $V_s/V_d$ ratio as well as the strength of the nematic order, which also sets the ratio between the $s$-wave and $d-$wave components similar to the previous sections. In other words, we argue the short time dynamics can be used in orthorhombic state to elucidate the nature of the primary and the secondary superconducting instabilities, possibly relevant to the case of FeSe as we discuss below. 

\begin{figure}
    \centering
    \includegraphics[width=\linewidth]{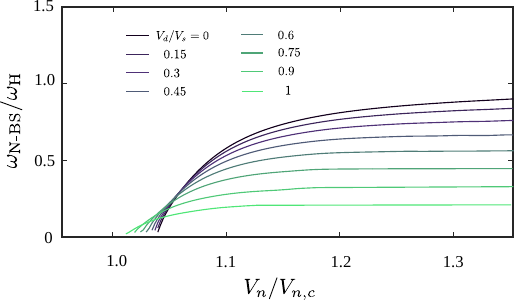}
    \caption{The nematic Bardasis-Schrieffer mode in the $s+d$-wave ground state depending on the strength of the nematic interaction $V_n$ for various ratios of $V_d/V_s$. The solid line is a solution of the linearized approximation in Eq.~(\ref{eq:lin_omega}). Due to the anisotropic order parameter the Bardasis-Schrieffer mode lies within the quasiparticle continuum for large enough $V_n$ and is then strongly dampened.}
    \label{fig:enhahncBS+nem}
\end{figure}

To get further insight into the collective modes in the mixed $s+d$ state we also compute their frequency dependence within linear response solving Eq.~(\ref{eq:lin_omega}). Similar to the $s$-wave ground state we find a (nematic) Bardasis-Schrieffer mode also inside the $s+d$-wave state and its frequency position is determined by the nematic interaction $V_n$ and the ratio $V_d/V_s$ as shown in Fig.~\ref{fig:enhahncBS+nem}. In particular, close to $V_{n,c}$, the mode appears to stiffen on increasing $V_n/V_{n,c}$, saturating to a finite value later. We attribute this to a hybridization of the amplitude mode of the nematic order and the BS mode; as the nematic mode becomes stiffer, the hybridization loses its importance. Next, we find that for a fixed value of $V_n > V_{n,c}$ the frequency of the in-gap mode strongly decreases as a function of $V_d/V_s$ but does not go to zero at $V_d/V_s = 1$. This decrease signals the transition from an $s$-wave driven superconducting $s+d$-state with $d$-wave as an nematicity induced byproduct into an $d$-wave driven $s+d$-state with $s$-wave as its an byproduct. However, this is not a thermodynamic phase transition, since no additional symmetries break, but a crossover and thus the in-gap mode can not go to zero at $V_d/V_s = 1$. Instead the frequency of this mode drops to a finite value. Additionally, the observability of this mode depends strongly on its location with respect to the onset of the quasiparticle continuum; e.g., as nematicity is enhanced, the gap develops deep minima and an enhanced damping of the nematic Bardasis-Schrieffer mode is expected, as is seen in Fig. 6. Note that the values of $V_n$ used here are sufficiently large, such that on increasing $V_d$, we avoid the $s+e^{i\alpha}d$ phase (see Fig. 1 (b)); on the boundaries of that phase a soft mode associated with time-reversal symmetry breaking would be expected\cite{kang2018}.

\section{Discussion and Conclusion}
One of the main conclusions of our study is that the nematic interaction in the system has a strong impact onto the short-time dynamics of unconventional superconductors. In particular, we argue that its presence  can yield additional oscillations in the time evolution of the superconducting  order parameter. These oscillations are in addition to the Higgs modes and appear as $B_{1g}$-symmetric Bardasis-Schrieffer type modes in the $s$-wave superconducting ground state, which appear (in the presence of nematic interactions) even if the interaction in the subdominant $d$-wave Cooper-pairing channel is weak. For the $d$-wave ground state, the nematic interactions yield additional weakly damped oscillations, whereas the pure s-wave BS modes are expected to be strongly damped by nodal quasiparticles. Furthermore, we argue that even if the tetragonal symmetry is explicitly broken in the nematic ground state the Higgs spectroscopy could be used to elucidate the nature of the primary and secondary superconducting instabilities, as the mode properties, such as lineshape, appear to be strongly dependent on the nature of the primary state. In this regard it is of interest to investigate further the short time dynamics of the iron-based superconductors, where both ingredients, strong competition between $s_\pm$- and $d$-wave superconductivity and nearby nematic order, are present and BS like mode has been observed \cite{Wu17,Boehm18}. We notice that actually both the nematic interaction, $V_n$, and the subleading $d$-wave interaction, $V_d$, would yield a single collective mode in the superconducting state of mixed origin. 
An additional interesting system to study would be FeSe$_{1-x}$Te$_x$ and FeSe$_{1-x}$S$_x$ compounds, as the parent  FeSe undergoes a nematic phase transition below 90 K and a superconducting transition below 8 K. The total $C_2$-symmetric superconducting order parameter is strongly anisotropic and (nearly) nodal. Substituting Se by S or Te induces an orthorhombic to tetragonal transition, yet superconductivity is more continuous function of $x$. Here a strong interplay of nematicity and various competing superconducting states would yield an interesting short-time dynamics. While we expect some quantitative difference due to the multi-band nature, the qualitiative picture of mixing of Bardasis-Schrieffer modes and nematic modes should not be affected. 

One other possible material class, where nematicity would affect the short time dynamics of the superconducting state  are the cuprates. Recent experiments found signatures of a second collective mode inside the superconducting phase of cuprates beside the Higgs-mode and its origin is still not clarified.\cite{Chu2020} As we argue in our paper the nematic mode, coupled to the Higgs oscillations of the $d-$wave superconducting ground state could be another possibility. In contrast to the Bardasis-Schrieffer mode, the nematic mode is much weaker damped by the nodal quasiparticles and can be a natural candidate for these additional oscillations.

In summary, we have studied the interplay between nematic order or fluctuations and Bardasis-Schrieffer modes in unconventional superconductors. We have shown that nematic order fluctuations and $B_{1g}$ symmetric Bardasis-Schrieffer mode mix into a single mode in the $s$-wave superconducting state, with its frequency softening in proximity to the boundary of the nematic and $s+id$ states, where the mixing is negligible. We have also shown that in the superconducting $d$-wave ground state the nematic interactions lead to a reduction of damping of the mixed BS-nematic mode. Finally, in the mixed state $s+d$ superconducting ground state, we have shown that lineshapes of the collective oscillations can be used to elucidate the nature of the primary and secondary superconducting instabilities.

\section{Acknowledgement}
P. A. V. acknowledges a Postdoctoral Fellowship from the Rutgers University Center for Materials Theory.

\appendix
\section{Collective modes with repulsive nematic interactions\label{app:B}}
\begin{figure}
    \centering
    \includegraphics[width=\linewidth]{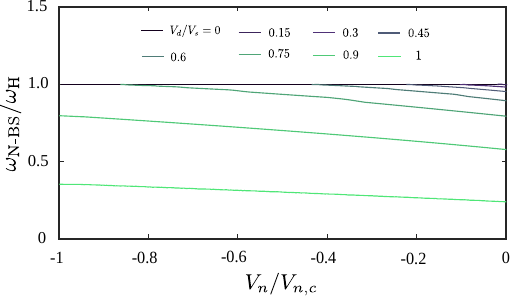}
    \caption{The dependence of the Bardasis-Schrieffer mode frequency $\omega_{\text{BS}}$ with respect to repulsive nematic interaction $V_n$ is computed from Eq.~(\ref{eq:lin_omega}) for various rations of $V_d/V_s$. At a certain strength of $V_n$ the Bardasis-Schrieffer mode is pushed into the Higgs-mode.}
    \label{fig:BS_vs_repnem}
\end{figure}
In this section we discuss the effect of repulsive nematic interactions onto the collective mode. This implies that $V_n > 0$ and therefore never leads to an emerging finite order parameter $\Delta_n$ and, thus, $C_4$-rotational symmetry is always present. \\In Fig.~\ref{fig:BS_vs_repnem} we present the solution of Eq.~(\ref{eq:lin_omega}) for $V_n >0$. Indeed we find that unlike attractive interactions, which support the Bardasis-Schrieffer mode by pushing it away from the quasiparticle continuum, the repulsive nematic interactions have the opposite effect and suppress the Bardasis-Schrieffer mode by pushing it closer to it. Depending on the exact ratio, the mode is pushed into the Higgs-mode frequncy and converges at some point with it. The larger the ratio $V_d/V_s$, the stronger is the repulsive nematic energy needed to push the mode into the continuum. The reason for that is not only because the Bardasis-Schrieffer mode frequency for small ratios $V_d/V_s$ is already close to the continuum but also the increase of the frequency seems to depend on this ratio, because for small ratios of $V_d/V_s$, this increase is stronger. However, independent of how close the $d$-wave state to the $s$-wave ground state is the Bardasis-Schrieffer mode is finally pushed into the Higgs-mode if the repulsive nematic interaction is strong enough. 

\section{Symmetry of nematicity\label{app:C}}
Here we show that a $B_{2g}$-nematic order does not interact with the Bardasis-Schrieffer mode of the $B_{1g}$ ($d_{x^2-y^2}$-wave) pairing channel. To do so, we analyze Eq.~(\ref{eq:delta_w}) but change the form factor of nematic order to the $B_{2g}$-symmetric form factor $\gamma_\kk^n = \sqrt{2}\sin(2\phi_\kk)$. We explicitly write the response matrix $\chi(\omega)$ for $\delta\boldsymbol{\Delta} = (\Delta_s^\prime,\Delta_s^{\prime\prime},\Delta_d^\prime,\Delta_d^{\prime\prime},\Delta_n)^T$
\begin{widetext}
\begin{align}
    &\chi(\omega) = 2\sum_\kk V G_\kk^T M_\kk G_\kk \nonumber \\
    &=\sum_\kk\frac{1}{E_\kk(\omega^2-4E_\kk^2)}\begin{pmatrix}
	&  &  \vdots&  & -2V_s \xi_\kk \Delta^\prime_\kk\gs\gn  \\ 
	&  &   \vdots& &  i V_s \omega\Delta^\prime_\kk\gs\gn \\ 
	\cdots&  \cdots&\chi_{s-d}^{4\times4} & \cdots  &  -2V_d \xi_\kk \Delta^\prime_\kk\gd\gn\\ 
	&  &  \vdots&  &  i V_d \omega\Delta^\prime_\kk\gd\gn\\  
	 4V_n \xi_\kk \Delta^\prime_\kk\gs\gn & -i 2V_n \omega\Delta^\prime_\kk\gs\gn & -4V_n \xi_\kk \Delta^\prime_\kk\gd\gn & -2i V_n \omega\Delta^\prime_\kk\gd\gn & 4V_n(\Delta^\prime_\kk)^2(\gn)^2
	\end{pmatrix} \label{eq:lin_orth}
\end{align}
\end{widetext}
The first four rows and columns connect real and imaginary parts of $\Delta_s$ with real and imaginary parts of $\Delta_d$. The fifth component of $\delta\boldsymbol{\Delta}$ is nematic order and thus the $(5,i)$- and $(i,5)$-component of the response function for $i\neq 5$ describes coupling of nematicity to superconductivity. From Eq.~(\ref{eq:lin_orth}) we find that if one assumes a $B_{2g}$ nematic form-factor $\gn = \sqrt{2}\sin(2\phi_\kk)$, then all of these $(5,i)$- and $(i,5)$-components vanish by symmetry because the $B_{2g}$-form factor is orthogonal to the $B_{1g}$ form factor of $d_{x^2-y^2}$-wave and $A_{1g}$ form factor of $s$-wave superconductivity. Thus the response function consists of two independent block matrices: a $4\times4$-matrix, describing the collective response of superconductivity and, thus, the Bardasis-Schrieffer and Higgs-modes and a $1\times 1$-matrix, describing the response of nematicity. Thus, the solution for the Bardasis-Schrieffer mode and the mode of the $B_{2g}$-nematicity in Eq.~(\ref{eq:lin_omega}) are independent of each other, showing that the coupling between nematicity and the Bardasis-Schrieffer mode is indeed only present if the nematic fluctuations and the subdominant pairing have the same symmetry.

\bibliography{bibliography.bib}
\end{document}